\documentclass{osa-article}
\pdfoutput=1
\usepackage{lineno}
\usepackage{breqn}
\usepackage{amsmath}
\usepackage{float}
\usepackage{gensymb}
\usepackage{textcomp}

\begin{document}
\textbf{PREPRINT}. \textit{Submitted for publication in Optics Express (2021)}

\title{Broadband-excitation-based mechanical spectroscopy of highly viscous tissue-mimicking phantoms}

\author{Magdalena A. Urbańska\authormark{1,2,*}, Sylwia M. Kolenderska, Sophia A. Rodrigues\authormark{3}, Sachin S. Thakur\authormark{4,5} and Frédérique Vanholsbeeck\authormark{1,2}}

\address{\authormark{1}The Dodd-Walls Centre for Photonic and Quantum Technologies, Department of Physics, University
of Auckland, New Zealand\\
\authormark{2}The Department of Physics, University of Auckland, Auckland 1010, New Zealand\\
\authormark{3}The Department of Chemical and Materials Engineering,  University of Auckland, New Zealand\\
\authormark{4}Buchanan Ocular Therapeutics Unit, Department of Ophthalmology, New Zealand National Eye Centre, Faculty of Medical and Health Sciences, University of Auckland,Private Bag 92019, Auckland 1142, New Zealand\\
\authormark{5}School of Pharmacy, Faculty of Medical and Health Sciences, University of Auckland, Private Bag 92019, Auckland 1142, New Zealand\\}
\email{\authormark{*}murb435@aucklanduni.ac.nz} 


\begin{abstract*}
Standard rheometers assess mechanical properties of viscoelastic samples up to 100~Hz, which often hinders the assessment of the local-scale dynamics. We demonstrate that high-frequency analysis can be achieved by inducing broadband Rayleigh waves and monitoring their media-dependent propagation using Optical Coherence Tomography. Here, we present a new broadband wave analysis based on two-dimensional Fourier Transformation. We validated this method by comparing the mechanical parameters to monochromatic excitation data and a standard oscillatory test. Our method allows for high-frequency mechanical spectroscopy, which could be used to investigate the local-scale dynamics of different biological tissues and the influence of diseases on their microstructure.
\end{abstract*}

%%%%%%%%%%%%%%%%%%%%%%%%%%  body  %%%%%%%%%%%%%%%%%%%%%%%%%%
\section{Introduction}

Extracting information about sample mechanical properties, such as elasticity and viscosity, has applications in industrial production and scientific research~\cite{Schroyen2020Bulk}. A form of mechanical spectroscopy that provides this information as a function of frequency is oscillatory rheometry. This method allows to assess fast dynamics and, thus, the local-scale microstructure of a sample when conducted at high frequencies~\cite{Schroyen2020Bulk,fritz2003characterizing}. The frequency range of this high-frequency regime depends on the medium and often exceeds the limitation of standard rheometers~\cite{willenbacher2007dynamics}. With  standard rheometers, high-frequency information about the sample behaviour beyond the limit of the device can be obtained using the temperature-time superposition principle~\cite{MalvernInstruments2016ARheology} which states that temperature and frequency can be treated as equivalent. However, this principle does not hold for many complex fluids and soft solids~\cite{Schroyen2020Bulk}. As a result, researchers have turned to alternative methods such as those based on excitation and detection of mechanical waves, which allow for high-frequency, non-invasive measurements. 

In the methods mentioned above, a stimulus is applied inside the sample or on its surface. This stimulus is a source of different waves, such as the surface waves: Rayleigh wave or Scholte wave propagating on the sample surface and the bulk waves: shear wave and compressional wave propagating inside the sample. Other waves that can also be observed in thin layered samples are guided Lamb waves. All these waves, except for the compressional wave, can be used to assess the sample mechanical properties by extracting information about their propagation parameters such as velocity and attenuation. The displacements of the sample structure caused by the propagation of these mechanical waves are observed inside the sample or on its surface by using different techniques such as laser vibrometry, ultrasound, magnetic resonance imaging or optical coherence tomography (OCT). Since OCT combines high-resolution, non-contact quantitative measurements with clinical applicability not achievable by other methods, its implementation to detect mechanical waves propagation is widespread. 

A great deal of research focused on monitoring the propagation of the mechanical waves using OCT assumes a purely elastic sample~\cite{Nguyen2015Shear,Lan2017Common-path,Song2013Shear,Ambrozinski2016Acoustic,Singh2018Quantifying,Qu2018In}. However, the results obtained with such an assumption have limited validity as tissue-like samples display not only elastic behaviour but also some degree of viscous behaviour. Therefore, researchers have started to study a more realistic, i.e. viscoelastic behaviour where velocity dispersion originating from the sample viscosity is considered. For this purpose, the frequency-dependent phase velocity values are extracted and fitted with different mathematical models whose frequency-independent parameters describe the elasticity and viscosity of a sample. Although this model-dependent approach often enables to describe a sample with two or three parameters allowing for a straightforward comparison between different samples, the viscoelastic parameters estimated using a specific viscoelastic model might be misleading as each model a priori assumes a particular behaviour of the sample. That assumption may be a source of misinterpretation of the sample viscoelasticity, especially for complex fluids and soft solids which express a complex behaviour. These complex samples can be described with complex models that use more parameters than standard models. However, the viscoelastic parameters obtained with these models are sensitive to small data fluctuations, which means these parameters are hard to reproduce~\cite{Klatt2007NoninvasiveViscoelasticity}.

N. Leartprapun~\textit{et al.}~\cite{Leartprapun2017Model-independent} presented a model-independent, mechanical, spectroscopic analysis using monochromatic stimuli. They induced a series of monochromatic shear waves in a sample with an acoustic radiation force and recorded the wave-induced displacements with an OCT system. By observing the phase shift of the wave and the decrease in amplitude with propagation distance, they obtained frequency-dependent information about the elasticity and viscosity of a sample in the form of the shear storage and loss moduli, respectively. These parameters allowed for a comprehensive analysis of the sample viscoelastic behaviour. However, the monochromatic signal cannot be generated for some types of stimulus source such as an air puff or laser pulse. For such excitation methods, broadband signals have to be analysed. Therefore, a model-independent technique that allows to use a broadband stimulus and extract the frequency-dependent viscoelastic properties of a sample is needed. In such a method, it is necessary to analyse the propagation of a broadband wave and extract information about its phase velocity and attenuation for all the temporal frequencies of the broadband signal.

Methods that allow to extract the information about the phase velocity for different frequencies progressed greatly through recent years. In 2004, S. Chen~\textit{et al.}~\cite{Chen2004Quantifying} proposed measuring the velocity dispersion curves by using multiple measurements with a monochromatic excitation and determining the phase velocity for each frequency detecting wave-induced displacement with a laser vibrometer at two lateral positions. Later, this method was improved upon by using Fourier Transformation (FT), which allowed for the analysis of broadband signals. The phase shift of the wave at each lateral position on the sample was extracted using 1D FT of the temporal displacement plots, and the phase velocity for each frequency of the broadband signal were calculated~\cite{Wang2014NoncontactCornea,Han2015Quantitative,Han2017Assessing}. Further, 2D FT of the displacement field graph, consisting of displacement plots for multiple lateral positions, was proposed to obtain the velocity dispersion curves for a broadband stimulus~\cite{Du2019Quantitative,Shih2018QuantitativeModel,Bernal2011MaterialModes,Couade2010QuantitativeImaging}. By using the absolute value of this 2D Fourier transform, a dispersion graph with the information about temporal and spatial frequencies of the mechanical wave was created. The temporal and spatial frequencies corresponding to the maximum amplitudes in the dispersion graph were extracted. Dividing one frequency by the other for each maximum provided the phase velocity of the wave at each particular temporal frequency of the broadband signal. It should be noted that in these studies, while the velocity dispersion was measured, different mathematical models were assumed to provide information about the viscoelasticity of the sample. Optimally, the attenuation also should be determined so the frequency-dependent viscoelastic parameters can be calculated using a model-independent approach.

Attenuation is rarely analysed for a broadband stimulus. However, there were a few interesting approaches to obtain information about the frequency-dependent value of the attenuation. A. Ramier~\textit{et al.}\cite{Ramier2019Measuring} extracted the frequency-dependent attenuation values using 1D FT along the temporal axis of the displacement field and fitted an exponential decay for each temporal frequency. Another method for the attenuation estimation of the mechanical waves was reported by I. Z. Nenadic \textit{et al.}\cite{Nenadic2017Attenuation}. In this method, a relationship between attenuation and the Full-Width Half-Maximum (FWHM) of the amplitude signal in the dispersion graph was derived. However, both these methods highly depend on the geometry of the wavefront and amplitude variations caused by the scattering attenuation~\cite{Quan1997SeismicMethod}. S. Bernard~\textit{et al.}~\cite{Bernard2017ATissuesc} proposed to use the frequency shift method developed in seismology~\cite{Quan1997SeismicMethod} instead to obtain the attenuation values, as this method is less dependant on the wavefront geometry or amplitude variation. However, a linear relationship between the attenuation and the temporal frequency is assumed. Moreover, for the attenuation measurements, as for the velocity, the model-dependent approach was chosen.

Here we present a novel model-independent approach allowing for mechanical spectroscopy with a broadband wave. We show that such an approach is especially beneficial for samples that express highly viscous behaviour. In this approach, we have focused on obtaining a reliable viscoelastic analysis of highly viscous samples, which requires obtaining the attenuation values independent of the wave amplitude. We demonstrate a quasi amplitude-independent method for the attenuation calculation by extracting the attenuation values using the imaginary part of the 2D FT of the 2D displacement field graph. Also, as most studies present just model-dependent solutions for assessing the viscoelastic parameters of a sample, we have implemented a model-independent method based on the frequency-dependent velocity and attenuation values of the wave to observe the frequency-dependent changes of the sample response. We have used viscoelastic samples which behaviour is expected to be strongly frequency-dependent, and we have discussed their relevance to the biological tissues.

In our measurements, we have induced Rayleigh waves with a piezoelectric transducer and observed their propagation with an OCT system. We have extracted their propagation parameters: velocity and attenuation. The velocity values were extracted from the dispersion graph and the attenuation values from the imaginary part of the 2D FT of the displacement field graph. Using the Rayleigh wave equation, these values were recalculated to the shear wave velocity and attenuation values. Based on these parameters, the shear storage modulus, shear loss modulus and phase angle between the viscous and elastic response were estimated, which allowed for a comprehensive analysis of the sample viscoelastic behaviour. First, a polydimethylsiloxane (PDMS) sample - a viscous soft tissue phantom - was used to validate our method for determining the shear moduli and the phase angle. It was done by comparing the values obtained for a broadband stimulus and multiple monochromatic stimuli. Next, the 2D FT method was used to provide the mechanical spectroscopy of a highly viscous sample, which has low elasticity, unlike the PDMS sample. For this purpose, we have used a phantom of an extracted vitreous humour exhibiting such behaviour. Since the measurements of this sample are time-sensitive due to its evaporation which alters its mechanical properties, the monochromatic measurements were not performed. Instead, for the vitreous humour phantom, the shear storage modulus, shear loss modulus and the phase angle values obtained using OCT (high frequency) were indirectly compared to data obtained with the rheological oscillatory test (low frequency).

\section{Materials and methods}

\subsection{Phantoms preparation}
A viscous tissue-mimicking phantom (PDMS, ELASTOSIL, RT 601 A/B) with a low proportion of a curing agent was used for the first experiment. This sample was prepared by thoroughly mixing components A and B in the proportion of 1:60. To eliminate bubbles in the sample, it was placed in a vacuum chamber for 30~min. Then, the sample was poured into a rectangular container of dimensions 14~mm x 30~mm x 50~mm and put for two hours in an oven set to 60°C. When the sample had cooled down, it was gently extracted from the mould.

For the second experiment, a vitreous humour phantom was made based on the guidelines presented in Ref.~\cite{Thakur2020ValidationEvaluations}. Two identical phantoms of the extracted vitreous humour were made by dissolving equal concentrations (0.75~mg/ml) of both agar and hyaluronic acid (HA -- hyaluronic acid sodium salt from Streptococcus Equi, Sigma-Aldrich, $1.5-1.8 x 10^6 $~Da) in saline prepared with one tablet of phosphate-buffered saline (Biotechnology Grade, VWR Life Science) in 100 ml of distilled water. The sample was made by adding 30mg of agar and 30mg of HA to 40ml of saline when heated to 100°C. Then, the solution was mixed thoroughly until no clumps were left and was left to cool down at room temperature~(21°C) until it reached equilibrium. Distilled water was added to compensate for any losses during heating and cooling. The sample was poured into a petri dish ($\varnothing$ 60~mm) to a height of approximately 10~mm. The measurements were done within a few hours of making the phantoms to limit any changes of the sample properties due to evaporation. One of the phantoms was used in the OCT measurement and the other in the oscillatory test.

\subsection{System description}
\begin{figure}[ht]
\centering\includegraphics[width=13.2cm]{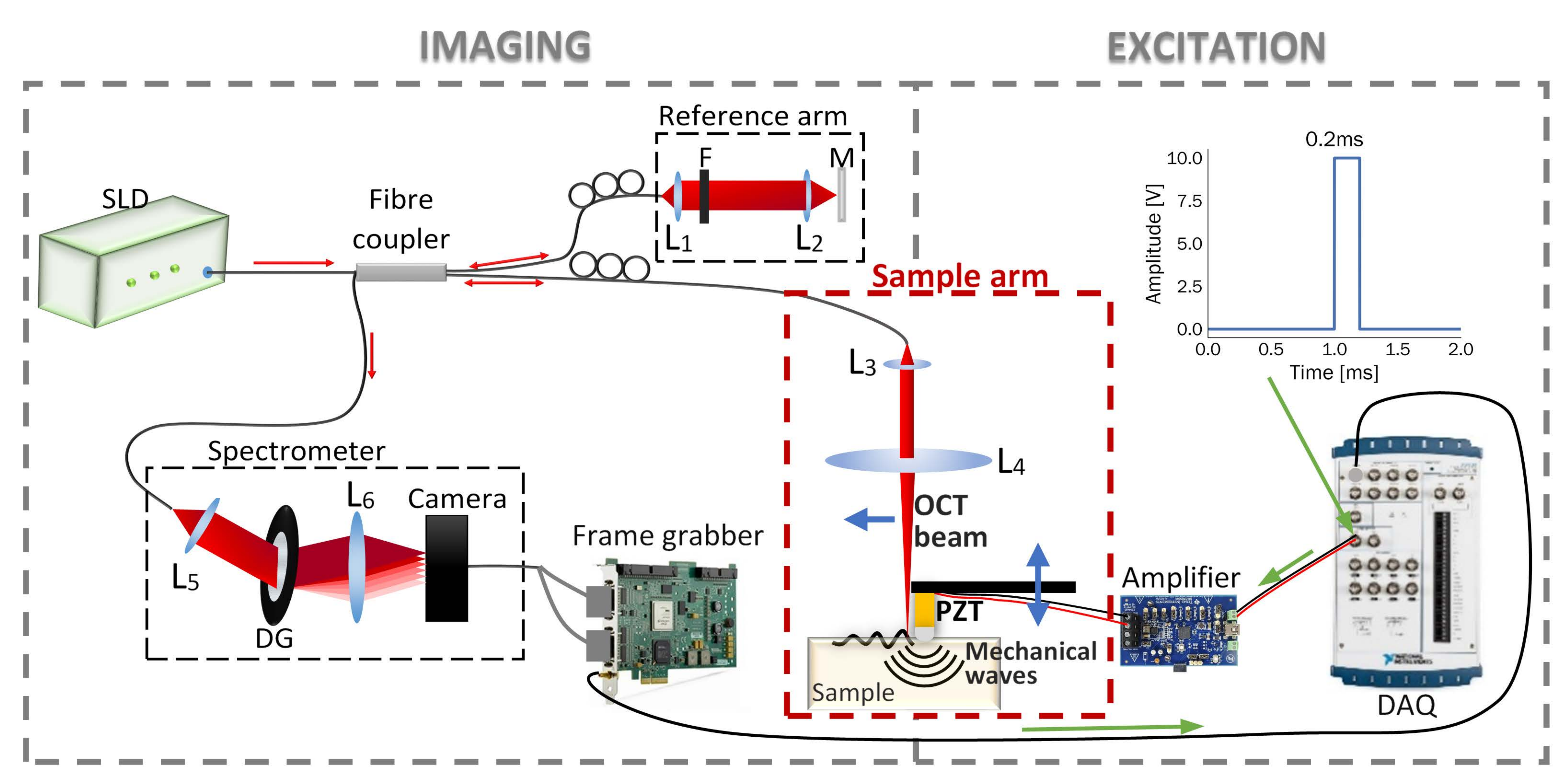}
\caption{The experimental system consists of the imaging part (marked with a dashed grey rectangle on the left) and the excitation part (marked with a dashed grey rectangle on the right). The excitation part, composed of a piezoelectric transducer (PZT), amplifier and a data acquisition (DAQ) card, is used to induce the mechanical waves in the sample. The imaging part, composed of an SD-OCT setup, consisting of a Michelson interferometer and a spectrometer, is used to observe the Rayleigh waves at the sample surface. The sample arm of the SD-OCT system is marked with a dashed red rectangle in the middle. The acquisition of the OCT camera triggers the signal presented at the right upper corner from the DAQ card. This signal is amplified to about 100V by the amplifier before the PZT. The red arrows indicate the direction of the optical signal, the green arrows the electrical signal and the blue arrow the mechanical movement. \mbox{L – lens}, \mbox{DG - diffraction} grating, \mbox{M – mirror}, \mbox{F - variable} neutral density filter, \mbox{S – scanning} mirror. Focal lengths of the lenses L1, L2, L3, L4, L5 and L6 are equal to 8.1~mm, 50~mm, 15~mm, 75~mm, 50~mm and 100~mm, respectively. The diffraction grating has 1200 lines/mm, and the camera has 8192 pixels (pixel size is 10~\textmu{}m).}
\label{fig:system}
\end{figure}

An SD-OCT system (Fig.\ref{fig:system}) was used to observe the wave-induced displacements. The light from a broadband source (SLD, Superlum Broadlighter T840, 780-920~nm) is directed to a Michelson interferometer with a 50/50 fibre coupler. The light from the interferometer is directed into the spectrometer with a 1200~lines/mm diffraction grating and a camera with a 70~kHz rate (spL8192-70km, Basler). The signal from the 2048 pixels of the camera covered by the spectrum is recorded using a frame grabber (PCIe-1433, National Instruments). The SD-OCT system has an axial resolution of $\sim$4~\textmu{}m and an imaging depth of $\sim$1.5~mm. Scanners were turned off during the measurements due to their low motion stability. Instead, a translation stage with 10~\textmu{}m precision is used to control the position of the OCT beam on the sample.

The mechanical waves are induced using a piezoelectric transducer (PZT) with an alumina hemispheric tip (2.5~mm tip diameter, PK4DLP1, Thorlabs). The PZT is attached to a mount and placed on a translation stage to bring it into physical contact with the sample. The displacement of the PZT towards the sample is kept below 3~µm. Before the measurement, the intensity of the backreflected light from the sample surface is maximised to ensure the direction of the beam incidence is perpendicular to the sample surface. The OCT beam is positioned as close as possible to the PZT tip without overlapping it.

\subsection{Data acquisition}
A program created in the LabVIEW environment is used to send the stimulus signal through the data acquisition (DAQ) card (NI 6251) to the PZT. The camera acquisition triggers a signal from the frame grabber to be sent to the DAQ card to synchronise the image acquisition with the stimulus. The stimulus commences 1~ms after the start of the acquisition. A voltage amplifier (DRV2700) is used to amplify the 10~V signal from the DAQ card to about 100~V to obtain adequate displacements of the PZT.
In the first experiment with the PDMS sample, two types of stimuli are used to excite the Rayleigh wave: a series of sinusoidal signals of different frequencies (0.2~kHz, 0.3125~kHz, 0.5~kHz, 1.0~kHz, 1.6~kHz) and a broadband stimulus (0.2~ms rectangular pulse). In the second experiment with the vitreous humour phantom, only the broadband stimulus is applied. After the stimulus is supplied to the sample, the information about the displacements at a given point on the sample surface is collected in the form of 2000 consecutive spectra. A set of 2000 spectra is collected at 49 and 15 specific lateral positions by gradually translating the OCT system's object arm by 0.2~mm and 0.1~mm,for the first and the second experiments, respectively. The smaller distance over which the measurements are done for the second sample is due to a lower uniformity in its surface flatness, which adds uncertainty to the beam-to-surface perpendicularity, and thus, adds uncertainty to the stimulus-to-detection distance for further distances on the sample. To improve the sensitivity of the measurement, 10 and 20 sets of data are obtained at each lateral position for the first and the second experiment, respectively. Each set is measured over a 28~ms long period.

\subsection{Data processing}
Each collected spectrum is Fourier transformed. The resulting complex-valued Fourier transform's polar components, amplitude and phase, provide the necessary information for analysis of the mechanical waves propagating through the sample. The amplitude is used to obtain the structural image of the sample and, therefore, the location of the sample surface peak. Then, the phase values corresponding to the pixels representing this peak are used to visualise the surface displacements. For each of these pixels, phase changes between the consecutive spectra in time are calculated, consequently generating phase-based displacement plots. After phase unwrapping, these displacement plots for all the chosen depths are averaged. The amplitude of the phase-based displacement plots is converted from the local phase difference, ${\Delta\phi(t)}$, between adjacent spectra, expressed in radians, to the amplitude of the displacement, ${u_z(t)}$, expressed in micrometres, using the formula:
\begin{equation}\label{eq:1}
u_z(t)=\frac{\Delta\phi(t)\lambda_c}{4\pi},
\end{equation}
where ${\lambda_c}$ is the central wavelength of the OCT system and t is time. There is no need to correct for the refractive index as just the displacement of the surface peak is observed since the sample is transparent. For the vitreous humour phantom, a moving average of width two is used to improve the signal to noise ratio of the displacement plots for each lateral position. This procedure is not necessary for the PDMS sample.

Different processing methods are used to obtain the velocity and attenuation of the Rayleigh waves for the data obtained with the monochromatic and broadband stimuli. For the monochromatic, or single tone, stimulus, the average displacement plots for consecutive lateral positions on the sample are cross-correlated, providing information about the time delay of the wave in these displacement plots. These time delays are plotted against the lateral positions and fitted linearly. The slope of the fit corresponds to the phase velocity of the wave propagating in the sample. The attenuation of these waves is calculated by fitting an exponential function to the wave amplitude-position relation and determining the damping parameter. The uncertainty of measurement for the velocity and attenuation is obtained by using the 95\% confidence interval. 

For the broadband stimulus, the displacement plots for multiple lateral positions on the sample are stacked to form the displacement field graph. 2D FT is applied to the displacement field graph to extract both the phase velocity and attenuation values. The phase velocity values are retrieved using the absolute value of the 2D Fourier transform. The resulting 2D array is called the dispersion graph and shows the spatial and temporal frequencies present in the displacement plots. In the dispersion graph, the spatial frequencies ($\xi$) corresponding to the maximum amplitudes for each temporal frequency ($f$) are extracted. These frequencies are used to calculate the phase velocity of the Rayleigh wave, ${c_R(f)}$:
\begin{equation}\label{eq:2}
	c_R(f)=\ \frac{f}{\xi(f)}.
\end{equation}
The information about the attenuation is obtained from the imaginary component of the 2D Fourier transform. The quarter corresponding to the negative spatial frequency and positive temporal frequency is chosen. The resulting 2D array, called the attenuation graph, presents the relationship between the negative spatial frequencies and the temporal frequencies. In the attenuation graph, the frequency-dependent attenuation values are obtained by determining the spatial frequencies corresponding to the maximum amplitude of the lobe traversing zero at each temporal frequency. The extracted spatial frequencies, ${\xi(f)}$, are used to obtain the attenuation values, ${\alpha\ (f)}$, where:
\begin{equation}\label{eq:3}
	\alpha\ (f)=-\xi(f).
\end{equation}
The uncertainty of measurement for the velocity and attenuation is calculated using the standard deviation. Since the size of the distance axis of the 2D displacement field graph consists of only 15 or 49 elements (depending on the measurement), as opposed to 2000 elements for the time axis, the distance axis is zero-padded, so its size is increased to 16,384 elements. This procedure is necessary to ensure optimum sampling on the spatial frequency axis of the 2D Fourier transform. Using the Rayleigh wave velocity and attenuation values, the shear wave velocity and attenuation values are calculated based on the Rayleigh wave equation.

\subsection{Rayleigh wave equation}
Based on the wave equation and the harmonic solution to this equation for an incompressible sample (compressional wave velocity $>>$ shear wave velocity), the Rayleigh wave equation describing the relation between the complex shear wave velocity, $c_s^\ast$, and complex Rayleigh wave velocity, $c_R^\ast$, can be expressed as~\cite{Carcione1992Rayleigh}:
\begin{equation}\label{eq:4}
\left(\frac{{c_R^\ast}^2}{{c_s^\ast}^2}\right)^3-8\left(\frac{{c_R^\ast}^2}{{c_s^\ast}^2}\right)^2+24\left(\frac{{c_R^\ast}^2}{{c_s^\ast}^2}\right)-16=0.
\end{equation}
This equation has real and complex solutions. The real solution is given as: 
\begin{equation}\label{eq:5}
\frac{c_R^\ast}{c_s^\ast}=0.955,
\end{equation}
and corresponds to the Rayleigh wave. The complex solution \cite{Schroder2001,Benech2017Analysis} is
\begin{equation}\label{eq:6}
\frac{c_R^\ast}{c_s^\ast}=1.97+i0.57,
\end{equation}
and corresponds to the leaky surface wave. Leaky surface waves have much higher velocity than the Rayleigh waves and small amplitudes that are often negligible~\cite{Benech2017Analysis}. Therefore, this solution will not be considered in further analysis.
The complex velocity in the Rayleigh wave equation can be expressed as~\cite{Carcione1992Rayleigh}:
\begin{equation}\label{eq:7}
c_m^\ast=\frac{\omega}{k_m^\ast},
\end{equation}
where ${k_m^\ast=\kappa_m-i\alpha_m}$, with $m:=R$ or $s$, is the complex wavenumber that describes both the wave dispersion with ${\kappa=\frac{\omega}{c_p}}$ where ${c_p}$ is the phase velocity, and the attenuation with ${\alpha}$ being the attenuation coefficient.
Based on Eq.(\ref{eq:5}) and Eq.(\ref{eq:7}), the relation between the wave numbers of the shear wave and Rayleigh wave can be expressed as:
\begin{equation}\label{eq:8}
k_s^\ast=0.95k_R^\ast,
\end{equation}
which can be then used to express the relation between the real and imaginary parts of these wavenumbers:
\begin{equation}\label{eq:9}
\kappa_s-i\alpha_s=0.95\kappa_R-i0.95\alpha_R,
\end{equation}
%\begin{equation}\label{eq:10}
%Re\left[k_s^\ast\right]=0.95Re\left[k_R^\ast\right]\ and\ Im\left[k_s^\ast\right]=0.95Im\left[k_R^\ast\right].
%\end{equation}
Thus, the relation between the velocities and attenuations of the shear and Rayleigh waves can be written as follows:
\begin{equation}\label{eq:11}
c_s=\frac{c_R}{0.95}\ and\ \alpha_s=0.95\alpha_R.
\end{equation}

Using Eq.(\ref{eq:11}), the Rayleigh wave velocity and attenuation are recalculated to the shear wave velocity and attenuation. Then, these values are used to estimate the frequency-dependent shear storage and loss moduli.

\subsection{Complex shear modulus}
The elastic and viscous behaviour of a tissue-like sample can be described with the complex shear modulus. This complex shear modulus, $G^{\ast}$, can be calculated using the complex wavenumber as follows:
\begin{equation}
\begin{split}
G^{\ast}\left(\omega\right)=&\rho\left(\frac{\omega}{k^\ast\left(\omega\right)}\right)^2=\rho\left(\frac{\omega}{\frac{\omega}{c_s\left(\omega\right)}-i\alpha_s\left(\omega\right)}\right)^2=\rho\left(\frac{\omega\left(\frac{\omega}{c_s\left(\omega\right)}+i\alpha_s\left(\omega\right)\right)}{\frac{\omega^2}{{c_s}^2\left(\omega\right)}+{\alpha_s}^2\left(\omega\right)}\right)^2=\\
=&\rho\frac{c_s^2\left(\omega\right)-\left(\frac{c_s^2\left(\omega\right)}{\omega}\right)^2\alpha_s^2\left(\omega\right)}{\left(1+\left(\frac{c_s\left(\omega\right)}{\omega}\right)^2\alpha_s^2\left(\omega\right)\right)^2}+i2\rho\frac{\frac{c_s^3\left(\omega\right)}{\omega}\alpha_s\left(\omega\right)}{\left(1+\left(\frac{c_s\left(\omega\right)}{\omega}\right)^2\alpha_s^2\left(\omega\right)\right)^2}=\\
=&G^\prime\left(\omega\right)+iG^{\prime\prime}\left(\omega\right)
\end{split}
\end{equation}
where $\omega$ is the temporal angular frequency, $\rho$ is the density of the sample, $G^\prime$ is the shear storage modulus and $G^{\prime\prime}$ is the shear loss modulus. Considering that we analyse our signal in temporal frequency, $f$, and assume only the positive values, the shear storage and loss moduli can be expressed as:
\begin{equation}\label{eq:G1}
G^\prime(f)=\left|\rho c_s^2(f)\frac{1-\left(\frac{\alpha(f)}{f}\right)^2c_s^2(f)}{\left(1+\left(\frac{\alpha(f)}{f}\right)^2c_s^2(f)\right)^2}\right|,
\end{equation}
\begin{equation}\label{eq:G2}
G^{\prime\prime}(f)=\frac{2\rho{\frac{\alpha(f)}{f}c}_s^3(f)}{\left(1+\left(\frac{\alpha(f)}{f}\right)^2c_s^2(f)\right)^2}.
\end{equation}

To express the behaviour of the sample with a single parameter, the phase angle between the elastic and viscous response, $\delta$, is obtained as follows:
\begin{equation}
\delta=arctan\left(\frac{G^{\prime\prime}}{G^\prime}\right).
\end{equation}
A phase angle equal to 0$^\circ$ and 90$^\circ$ describes a purely elastic and a purely viscous behaviour, respectively. For the single tone data, the uncertainty of measurement for shear moduli and the phase angle is calculated using the exact differential. For the broadband stimulus data, these uncertainties are calculated using the standard deviation.

\subsection{Oscillatory rheometry}
Oscillatory rheometry is used to indirectly validate the mechanical parameters obtained with the OCT system for the vitreous humour phantom. Therefore, an oscillatory frequency sweep test is done on the vitreous humour phantom using a controlled-stress rheometer (TA Instruments, Model ARG2) with a 1~mm gap between smooth, parallel plates of 40~mm diameter. For this measurement, the sample is inserted between the plates where the bottom plate is stationary, and the upper plate is rotated by a fixed small angle with increasing frequency (0.1-100 Hz). The values for frequencies above 50.12Hz were mostly negative. Since negative
values have no physical sense, they were treated as an error and only values obtained up to
50.12Hz were considered. The temperature for this test was set to 21ºC. The chosen fixed angle for this measurement corresponded to 1\% strain (average displacement 5~\textmu{}m). Amplitude sweeps were conducted to confirm that 1\% strain is well within the linear viscoelastic region. Additionally, this small strain corresponds to the strain used in rheological measurements of the vitreous humour~\cite{Sharif-Kashani2011RheologyProperties,Filas2014EnzymaticBody,Shafaie2018DiffusionVitreous} and can be compared to the stain induced by small displacements used in the OCT measurements. The measurements are repeated four times, and the measurement uncertainty is obtained using the standard deviation and t-distribution.

\section{Results and discussion}
\subsection{Viscoelastic analysis of the PDMS sample}
\begin{figure}[H]
\centering\includegraphics[width=13.2cm]{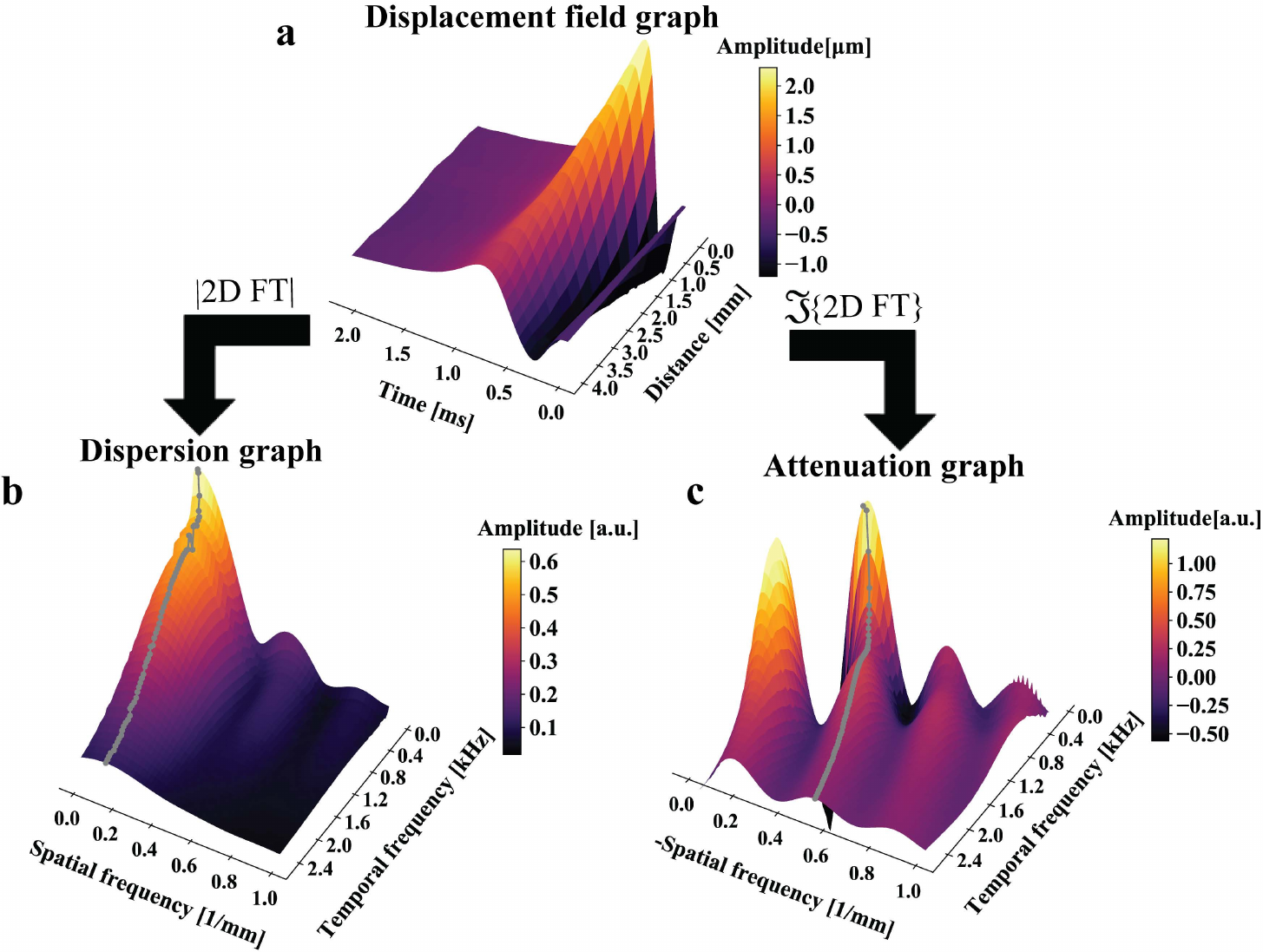}
\caption{The process of extracting velocity and attenuation values for the PMDS sample: displacement field graph presenting the propagation of the Rayleigh wave along a distance of 4~mm on the sample surface (a), dispersion graph based on the absolute value of the 2D Fourier transform of the displacement field graph with the maximum amplitude values of the main lobe marked with grey dots (b) and attenuation graph created using the imaginary part of the 2D Fourier transform with marked maxima for the lobe closest to traversing zero frequency (c).}
\label{fig:3Dgraphs}
\end{figure}

\begin{figure}[H]
\centering\includegraphics[width=13.2cm]{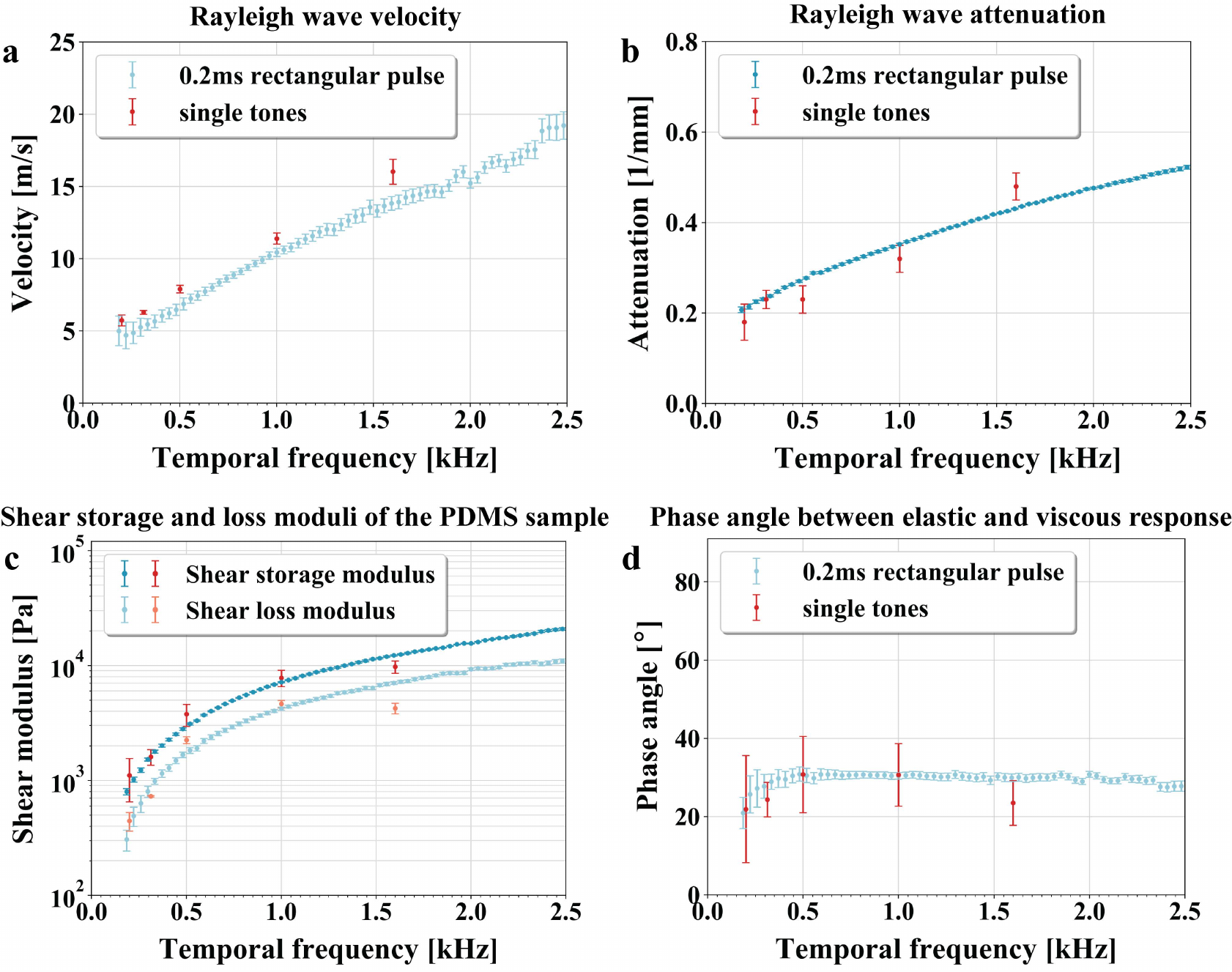}
\caption{The comparison of the data extracted for the single tone (red) and broadband (blue) stimuli for the PDMS sample. The Rayleigh wave velocities extracted from the dispersion graph for the broadband stimulus (blue) and for the discrete data based on the cross-correlation for the single tone stimuli (red) (a). The Rayleigh wave attenuation extracted from the attenuation graph for the broadband stimulus (blue) and the amplitude decay for the single tone stimuli (red) (b). The shear storage and loss moduli obtained using Eq.(\ref{eq:G1}) and Eq.(\ref{eq:G2}) for the broadband stimulus (blue) and the single tone stimuli (red) (c). The phase angle between the viscous and elastic response for the PDMS sample obtained using the broadband stimulus (blue) and the monochromatic stimuli (red) (d).}
\label{fig:graphs}
\end{figure}

The dispersion and attenuation graphs obtained for the PDMS sample are presented in Fig.\ref{fig:3Dgraphs}. In the dispersion graph (Fig.\ref{fig:3Dgraphs}b), a very clear main lobe with an additional side lobe, resulting from the shape of the signal, is observed. The attenuation graph consists of rather a periodic signal. The grey dots on the graphs correspond to the maximum amplitudes for which the coordinates were extracted. The data from the dispersion graph were used to calculate the phase velocity values using Eq.(\ref{eq:2}), and the data from the attenuation graph provided the information about the attenuation values based on Eq.(\ref{eq:3}). These values are presented in Fig.\ref{fig:graphs}a,b. 

As it can be seen in Fig.\ref{fig:graphs}~a and b, there is an approximately fourfold increase in the Rayleigh wave velocity over the presented frequency range (0.2-2.5kHz) and 2.5 times increase in the attenuation value. This high increase in the velocity is caused by the viscosity of the sample. The velocity values we obtained reflect those for cervical tissue~\cite{Torres2021ShearModel,Rus2020WhyDiagnosis} showing that this phantom is a good representation of a viscous tissue with a considerable elastic response. Cervical tissue has a high hyaluronic acid content, which plays a key role in the hydration of the tissue and makes it more viscous. To the best of our knowledge, the attenuation was not calculated for the cervical tissue; thus, this parameter cannot be compared.

To validate the velocity and attenuation values obtained with a broadband stimulus, we have compared them to the values obtained for several discrete frequencies (see Fig.\ref{fig:graphs}). The comparison shows a fair agreement between both methods. The slight disparity in the velocity data (Fig.\ref{fig:graphs}~a) can be due to not ideally monochromatic signals, which are difficult to obtain in practice.

Fig.\ref{fig:graphs}~c presents the frequency-dependent shear storage and loss moduli values. In this graph (Fig.\ref{fig:graphs}~), the curves for the shear storage and loss moduli are parallel to each other. The phase angle (Fig.\ref{fig:graphs}d) which expresses the relationship between the shear storage and loss moduli is approximately constant with frequency (Fig.\ref{fig:graphs}d). Both the shear moduli and the phase angle for the broadband stimulus are in the good agreement with the single tone data.

The phase angle allows using just one parameter to describe the changes in sample viscoelastic behaviour with frequency and helps to classify this behaviour. As the phase angle values considerably greater than zero suggest, the PDMS sample used in our experiments features a significant viscous response (shear loss modulus). However, it does not overcome the elastic response (shear storage modulus) as the phase angle is below 45$^{\circ}$ (Fig.\ref{fig:graphs}c). This type of behaviour can be classified as a gel-like viscoelastic solid behaviour. This behaviour of PDMS samples was also observed at low frequencies\cite{Valentin2019Substrate}, showing that this type of sample potentially maintains similar gel-like properties over a wide frequency range. 

\subsection{Viscoelastic analysis of the vitreous humour phantom}

\begin{figure}[H]
\centering\includegraphics[width=13.2cm]{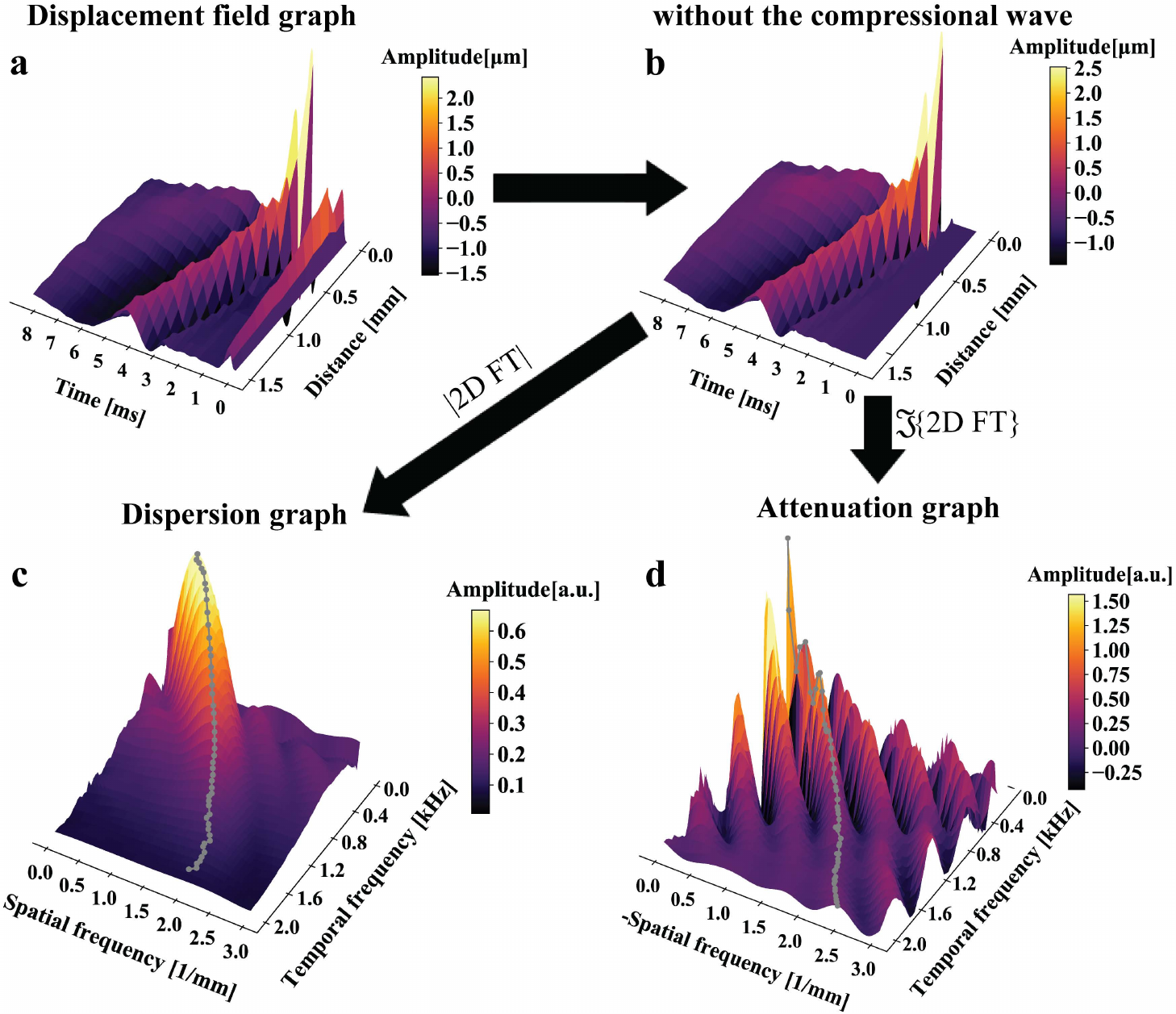}
\caption{The process of extracting velocity and attenuation values for the vitreous humour phantom: a displacement field graph presenting the propagation of the Rayleigh and compressional waves (a), displacement field graph presenting the propagation of only the Rayleigh wave through 1.5~mm on the sample surface (b), dispersion graph based on an absolute value of the 2D Fourier transform of the displacement field graph with the maximum amplitude values corresponding to the main lobe of the frequency signal marked with grey dots (c) and attenuation graph created from the imaginary part of the 2D Fourier transform with marked maxima for the lobe closest to traversing zero frequency (d).}
\label{fig:3Dgraphs0.75}
\end{figure}

The procedure of extracting the phase velocity and attenuation values for the vitreous humour phantom is as the one presented for the PDMS sample in Fig.\ref{fig:3Dgraphs} with the addition of two steps. The information about the displacement of the compressional wave~(Fig.\ref{fig:3Dgraphs0.75}a) was removed so that it does not alter the values extracted for the Rayleigh wave~(Fig.\ref{fig:3Dgraphs0.75}b), and a second-order Butterworth filter was used to improve the signal-to-noise ratio. The high amplitude of the displacement caused by the compressional wave is due to lower compressibility and lower thickness of the vitreous humour phantom compared to the PDMS sample.

\begin{figure}[H]
\centering\includegraphics[width=13.2cm]{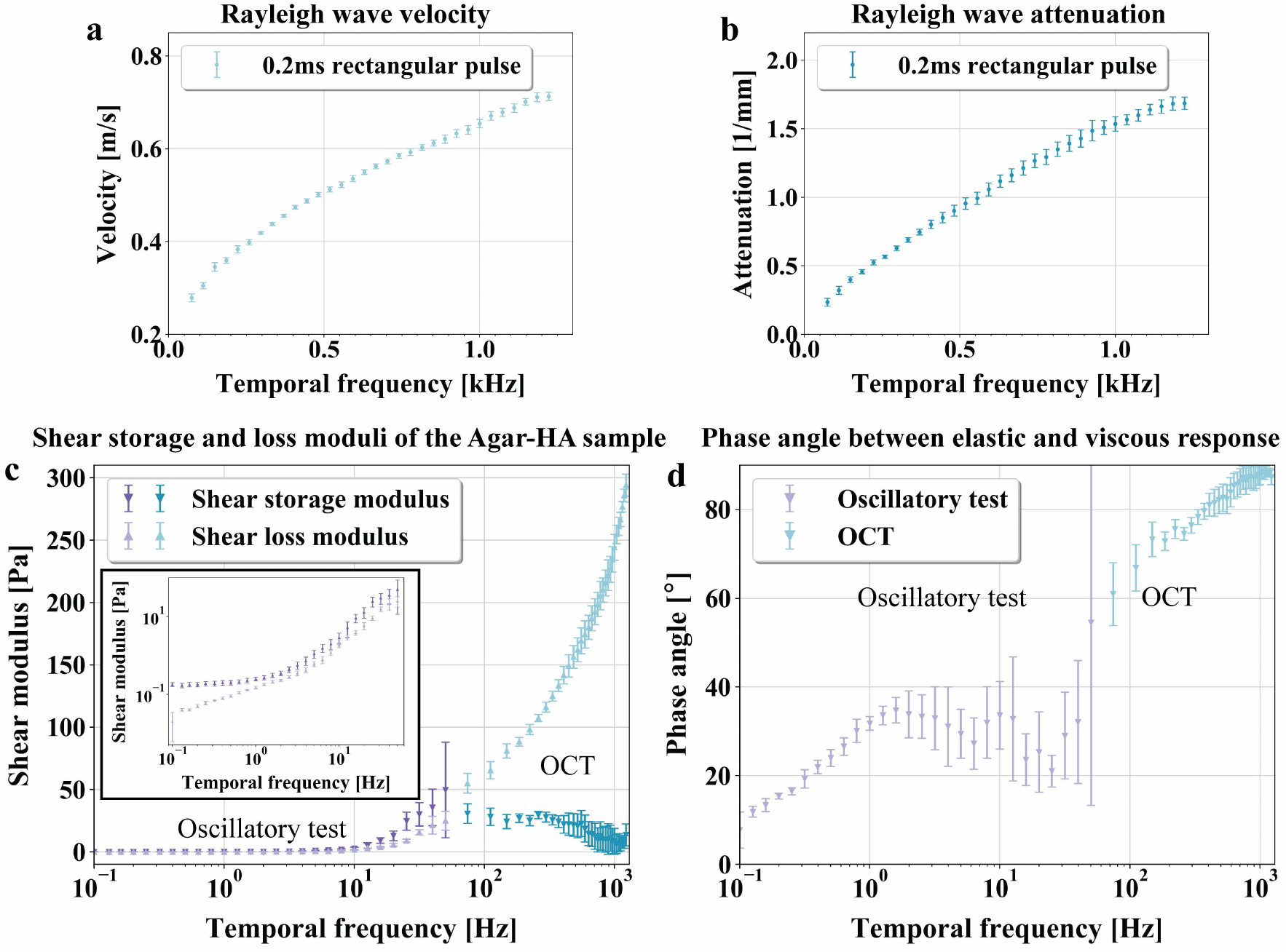}
\caption{The Rayleigh wave velocities extracted from the dispersion graph for the broadband stimulus (a). The Rayleigh wave attenuation extracted from the attenuation graph for the broadband stimulus (b). The shear storage and loss moduli obtained using Eq.(\ref{eq:G1}) and Eq.(\ref{eq:G2}) for the broadband stimulus (blue) and the values collected using oscillatory test (purple) (c). The phase angle between the viscous and elastic response for the vitreous humour phantom obtained using the broadband stimulus (blue) and oscillatory test (purple) (d).}
\label{fig:graphs0.75}
\end{figure}

The dispersion and attenuation graphs presented in Fig.\ref{fig:3Dgraphs0.75}~c and d, respectively, were generated in the same way as for the PDMS sample. As previously, the grey dots mark the maximum amplitudes for which the spatial and temporal frequencies were extracted. It can be observed that the spatial frequency values corresponding to the maximum amplitude in the dispersion graph (Fig.\ref{fig:graphs}c) increase much faster with the temporal frequency in comparison to the previous experiment. As previously, the velocity and attenuation of the Rayleigh wave are obtained using Eq.(\ref{eq:2}) and Eq.(\ref{eq:3}), respectively (Fig.\ref{fig:graphs0.75}). It is important to note that these parameters for the vitreous humour phantom are very different to those for other parts of the eye. For example, for the cornea which is more elastic and less viscous than the vitreous humour, velocity and attenuation values are about five times higher and five times lower, respectively~\cite{Ramier2019Measuring}. 

Since the OCT and rheometer measurements allowed us to extract the mechanical properties of the sample within different frequency ranges, the validation of the mechanical properties obtained with OCT for the vitreous humour phantom is not straightforward. As an indirect assessment, we have compared a transition between the high-frequency OCT and low-frequency rheological oscillatory test data. A seamless transition between these two data sets is evident (Fig.\ref{fig:graphs0.75}~c and d), supporting the reliability of the OCT approach. Additionally, the analysis of the results from these two tests allowed us to observe interesting changes of the vitreous humour phantom viscoelastic characteristics with frequency. Based on the results presented in Fig.\ref{fig:graphs0.75}~c and d, it can be stated that the sample is expressing elastic-dominant behaviour at low frequencies (the phase angle is below 45$^{\circ}$) and viscous-dominant behaviour at high frequencies (the phase angle is above 45$^{\circ}$). The transition between these two behaviours can sometimes be observed for entangled polymers~\cite{Duffy2016AnMode}, as the microstructure of the sample rearranges, and the stored elastic stresses relax and convert into viscous stresses \cite{MalvernInstruments2016ARheology}. The observed transition shows how using high frequencies provides us with additional information about the dynamics in the sample microstructure than can not be obtained from the low frequency data. Our noninvasive method is advantageous over photonic force optical coherence elastography (PF-OCE)~\cite{Lin2019Spectroscopic} which also allowed to observe such a transition for a hydrogel sample similar to our vitreous humour phantom but requires injection of polystyrene beads.

\section{Conclusion}
We have demonstrated that our model-independent method based on the 2D Fourier Transformation of OCT data for a broadband stimulus enables a high-frequency analysis of three standard rheometric parameters which describe the sample viscoelastic behaviour: the shear storage modulus, the shear loss modulus and the phase angle. Thus, this 2D FT method proves helpful as high-frequency rheometry. The comparison to single tone measurements and the oscillatory test validated the viscoelastic properties obtained with the 2D FT method. The advantage of this 2D FT model-independent analysis over the single tone measurements is that velocity and attenuation can be extracted for a broadband signal making it much more time-efficient. Additionally, it allows for different types of stimuli to be supplied to the sample, which cannot provide a monochromatic excitation such as an air puff or a laser pulse, improving upon the multiple single tone measurements presented by Lin Y.~\textit{et al.}\cite{Leartprapun2017Model-independent}. The obtained high-frequency information about the sample behaviour shows to be especially beneficial for highly viscous samples with low elasticity as their behaviour change notably for high frequencies where the effect of the fast dynamics of the local-scale microstructure can be observed.

\begin{backmatter}
\bmsection{Funding} The Marsden Fund from the Royal Society of New Zealand- Te Apārangi (UoA1509).

\bmsection{Acknowledgments}

\bmsection{Disclosures}
The authors declare no conflicts of interest.

\bmsection{Data Availability Statement}
Data underlying the results presented in this paper are not publicly available at this time but may be obtained from the authors upon reasonable request.

\end{backmatter}

%%%%%%%%%%%%%%%%%%%%%%% References %%%%%%%%%%%%%%%%%%%%%%%%%

\bibliography{ref-extracts}

\end{document}